\documentclass[journal=jpcafh,manuscript=article]{achemso}

\usepackage{rotating}
\usepackage{chemformula} % Formula subscripts using \ch{}
\usepackage[T1]{fontenc} % Use modern font encodings
\usepackage{amsmath}
\usepackage{graphicx}% Include figure files
\usepackage{bm}% bold math
\author{Karolina Janicka}
\author{Aleksander L. Wysocki}
\author{Kyungwha Park}
\affiliation[Virginia Tech]
{Department of Physics, Virginia Tech, Blacksburg, Virginia, 24061 USA}
\email{kyungwha@vt.edu}
\title[An \textsf{achemso} demo]
  {Computational Insights into Electronic Excitations, Spin-Orbit Coupling Effects,
  and Spin Decoherence in Cr(IV)-based Molecular Qubits}


\begin{document}

%\begin{tocentry}
%\centering
%\includegraphics[width=0.99\linewidth]{TOC.pdf}
%\end{tocentry}

%Manually done TOC Entry
%\AtEndDocument{
%\newpage
%\begingroup
%\section*{TOC Graphic}
%\sffamily
%\singlespacing
%\begin{center}
%\fbox{
%\begin{minipage}{3.25in}
%\vbox to 1.75in{\includegraphics{TOC}}
%\end{minipage}
%}
%\end{center}
%\endgroup
%}

\begin{abstract}
%250 words limit: ACS Nano
The great success of point defects and dopants in semiconductors for quantum information processing has invigorated a search for molecules with analogous properties. Flexibility and tunability of desired properties in a large chemical space have great advantages over solid-state systems. The properties analogous to point defects were demonstrated in Cr(IV)-based molecular family, Cr(IV)(aryl)$_4$, where the electronic spin states were optically initialized, read out, and controlled. Despite this kick-start, there is still a large room for enhancing properties crucial for molecular qubits. Here we provide computational insights into key properties of the Cr(IV)-based molecules aimed at assisting chemical design of efficient molecular qubits. Using the multireference {\it ab-initio} methods, we investigate the electronic states of Cr(IV)(aryl)$_4$ molecules with slightly different ligands, showing that the zero-phonon line energies agree with the experiment, and that the excited spin-triplet and spin-singlet states are highly sensitive to small chemical perturbations. By adding spin-orbit interaction, we find that the sign of the uniaxial zero-field splitting (ZFS) parameter is negative for all considered molecules, and discuss optically-induced spin initialization via non-radiative intersystem crossing.  We quantify (super)hyperfine coupling to the $^{53}$Cr nuclear spin and to the $^{13}$C and $^1$H nuclear spins, and we discuss electron spin decoherence. We show that the splitting or broadening of the electronic spin sub-levels due to superhyperfine interaction with $^1$H nuclear spins decreases by an order of magnitude when the molecules have a substantial transverse ZFS parameter.
\end{abstract}


\section{Introduction}
Point defect centers and dopants in wide-bandgap semiconductors have been shown to be viable options for quantum information processing \cite{Awschalom2018,Hensen2015,Taminiau2014,Epstein2005,Balas2008,Bassett2019,Fuchs2009,He2019,Fricke2021,Madzik2022}. Two representative examples are negatively charged nitrogen-vacancy (NV$^{-}$) centers in diamond \cite{Davies1976,Epstein2005,Rogers2008,Batalov2008}, and phosphorus dopants in silicon.\cite{He2019,Fricke2021,Madzik2022,Kane1998} Electronic spin states of the NV$^{-}$ centers can be optically initialized, read-out, and coherently controlled with long spin coherence time.\cite{ Awschalom2018,Hensen2015,Taminiau2014,Epstein2005,Balas2008,Bassett2019,Fuchs2009} Nuclear spin states of phosphorus ($^{31}$P) dopants can be controlled by a gate voltage or electric field.\cite{Kane1998} Both systems have been experimentally shown to perform quantum gate operations with high fidelity.\cite{Awschalom2018,Hensen2015,Taminiau2014,Bassett2019,Fuchs2009,He2019,Fricke2021,Madzik2022}

Inspired by this great success, organic radicals and transition-metal-based (TM) and lanthanide-based (Ln) magnetic molecules have been tailored to have desirable properties for quantum information science applications by utilizing the versatility of chemical environment.\cite{Rugg2019,Zadrozny2014,Atzori2016,Thiele2014,Godfrin2017,Shiddiq2016,Liu2021,Serrano2022,Fataftah2020,Wojnar2020} Either molecular electronic spin states or electronic-nuclear spin states can be considered as quantum bits (qubits) or quantum $d$-levels (qudits) which may be initialized, read-out, or controlled by an external magnetic field and/or electric field, or optical means. For organic donor-acceptor-radical molecules, the electronic spin states were shown to be entangled and teleported with high fidelity by using microwave pulses and photo-excitation.\cite{Rugg2019} As long as TM-based or Ln-based magnetic molecules are concerned, the hyperfine interaction between the molecular electronic spin and the TM (or Ln) nuclear spin has been mostly utilized to propose nuclear spin qubits.\cite{Thiele2014,Godfrin2017,Shiddiq2016,Liu2021} For vanadium(IV)-based magnetic molecules, the molecular electronic-nuclear spin states were shown to have long spin coherence time.\cite{Zadrozny2014,Atzori2016} For terbium(III)-based molecules, the electronic-nuclear spin states were shown to be initialized and read-out by an external magnetic field and to be manipulated by a gate voltage within a single-molecule transistor set-up,\cite{Thiele2014} where Grover's algorithm was also implemented.\cite{Godfrin2017} For holmium(III)-based molecules, the electronic-nuclear spin states were shown to undergo a clock transition (i.e., the level separations being insensitive to a magnetic field to first order)\cite{Shiddiq2016} and to be coherently controlled by distortion and an electric field.\cite{Liu2021} For europium(III)-based molecules, the nuclear spin states were optically initialized and controlled.\cite{Serrano2022}

Although electronic-nuclear or nuclear spin states are less susceptible to environment than electronic spin states, control and gate operations of the former are much slower than those of the latter. It is known that $4f$ orbitals of Ln-based molecules are typically highly localized well below the highest occupied molecular orbital, which hinders optical excitations of electronic states. Therefore, there may be a higher probability to find optically accessible molecules from TM-based molecular families,\cite{Fataftah2020,Wojnar2020,Bayliss2020} as far as electronic spin states are targeted. Along this line, a molecular analogue closest to the NV$^{-}$ center was discovered in a Cr(IV)(aryl)$_4$ family.\cite{Bayliss2020} The experiment\cite{Bayliss2020} showed that the electronic ground state of the Cr(IV)(aryl)$_4$ molecules has a spin $S=1$ (triplet) with a small zero-field splitting (ZFS) and that the electronic first excited state has a spin $S=0$ (singlet), which is the same as that of the NV$^{-}$ center in diamond. The molecular electronic states were optically initialized, read-out and coherently controlled with microwaves.\cite{Bayliss2020} Furthermore, ZFS parameters were shown to vary with modifications of ligands in the family,\cite{Bayliss2020,Laorenza2021} although the sign of ZFS parameter $D$ was not determined in the experiments.

Despite this progress, quantitative computational insights into the systems would facilitate the design of molecular qubits with enhanced properties such as narrower zero-phonon line (ZPL), more efficient spin initialization (or larger optical spin polarization), reduced intermolecular interactions, longer spin-lattice relaxation time, and longer spin coherence times. Multireference nature of electronic excitations plays a key role in absorption and emission properties, non-radiative transitions between the states, and ZFS (or magnetic anisotropy). An understanding of these properties under different chemical environment is crucial in increasing the optical spin polarization and elucidation and reduction of spin decoherence factors. So far, there have been no studies of these properties for the Cr(IV)(aryl)$_4$ family.

\begin{figure}[htb!]
\centering
\includegraphics[width=1.0\linewidth]{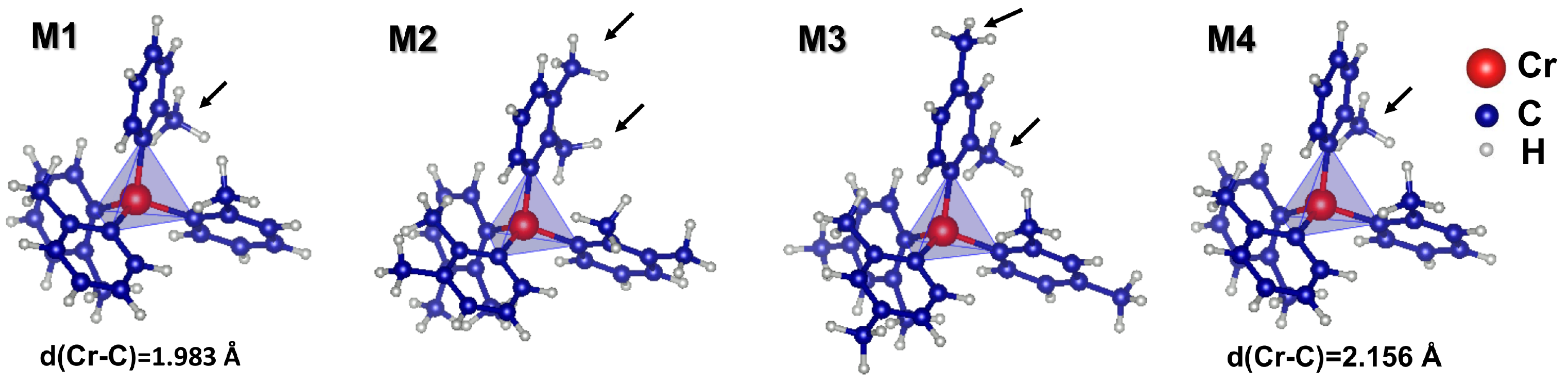}
\caption{Atomic structures of four Cr(IV)(aryl)$_4$ molecules: \textbf{M1}, \textbf{M2}, \textbf{M3} and \textbf{M4}. Red, dark blue and grey circles denote Cr, C and H atoms, respectively. In each molecule, an approximately tetrahedral C cage containing the Cr atom is shown as a light blue polyhedron. \textbf{M4} has the same molecular symmetry as \textbf{M1}. In the experimental geometries, {\bf M1} and {\bf M4} have different bond lengths and bond angles (see Table~\ref{tab:1} and main text for detail). Upon geometry relaxation, the atomic coordinates of {\bf M4} become the
same as those of {\bf M1}.}
\label{fig:geo}
\end{figure}

In this work, we investigate multireference electronic excitations of several Cr(IV)(aryl)$_4$ molecules (aryl$=$o-tolyl ({\bf M1},{\bf M4}, {\bf M5}), 2,3-dimethylphenyl {\bf M2}, 2,4-dimethylphenyl ({\bf M3})), as shown in Fig.~\ref{fig:geo}, using the multireference {\it ab initio} methods including spin-orbit coupling (SOC), and analyze the effect of different chemical environment on the excitations. Based on the calculated electronic spin-triplet and spin-singlet excitations, we study two SOC effects for the molecules such as (i) the ZFS of the ground spin-triplet state and (ii) transitions between spin-triplet and spin-singlet states (intersystem crossing, ISC), within the multireference {\it ab initio} formalism. Then we quantify hyperfine coupling to the $^{53}$Cr nuclear spin and to the $^{1}$H and $^{13}$C nuclear spins of the ligands, and discuss their effects on decoherence of the molecular electronic spin states for the different molecules.

\begin{table}[htb!]
\centering
\caption{Structural properties of all considered molecules (Fig.~\ref{fig:geo}) where the atomic coordinates of {\bf M1}, {\bf M2}, {\bf M3}, and {\bf M5} are taken from the experimental Cr(IV)(aryl)$_4$ molecular crystals,\cite{Bayliss2020} while those of {\bf M4} are from one of the experimental diluted molecular crystals with Sn replaced by Cr.\cite{Bayliss2020}}
\begin{tabular}{c|c|c|c|c|c}
\hline \hline
Properties                    & \textbf{M1}  & \textbf{M2}    & \textbf{M3}    & \textbf{M4}  & \textbf{M5} \\
\hline
Cr-C bond lengths (\AA)~      & 1.983        & 2.002, 2.017   & 1.990, 1.991   & 2.156        & 1.983, 1.992 \\
                              &              &                & 1.996, 1.998   &              &  1.992, 1.995 \\
C-Cr-C angle (degree)         & 113.07       & 109.55, 110.38 & 109.01, 109.34 & 107.14       & 113.28, 113.51 \\
                              &              &                & 109.72, 110.64 &              & 105.06, 102.37 \\
Molecular symmetry            & S$_4$        &  C$_2$         & C$_1$          & S$_4$        & C$_1$ \\ 
\hline
\hline
\end{tabular}
\label{tab:1}
\end{table}

All five molecules consist of a Cr$^{4+}$ ion in an approximately tetrahedral ($T_d$) environment produced by four surrounding aromatic hydrocarbon ligand rings. Each ligand ring corresponds to the benzene molecule with one or two hydrogen atoms being replaced by a methyl group. The \textbf{M1}, \textbf{M2}, and \textbf{M3} molecules differ by number and/or positions of the methyl group in the ligand rings (marked as arrows in Fig.~\ref{fig:geo}). Table~\ref{tab:1} lists exact molecular symmetry and bond lengths and angles of the Cr ion and the four closest C sites for the molecules with the experimental geometries.\cite{Bayliss2020} Note that both \textbf{M1} and \textbf{M4} molecules have exact $S_4$ symmetry but the Cr-C bond length differs by 0.173~\AA. The atomic coordinates of the {\bf M1}, {\bf M2}, {\bf M3}, and {\bf M5} molecules are taken from the experimental Cr(IV)(aryl)$_4$ molecular crystals, while those of the {\bf M4} molecule are from one of the synthesized diluted molecular crystals with a Cr:Sn ratio of 0.75\%.\cite{Bayliss2020} In this diluted crystal, since only atomic coordinates of a Sn(IV)(aryl)$_4$ cluster were reported,\cite{Bayliss2020} the {\bf M4} molecule is constructed by replacing Sn by Cr in the Sn(IV) cluster. Thus, the {\bf M4} molecular structure corresponds predominantly to the Sn(aryl)$_4$ geometry, and the calculations for the {\bf M1} molecule (rather than for {\bf M4}), therefore, better represent the experimental data for the $S_4$-symmetric Cr(IV)({\it o}-tolyl)$_4$ molecule. Nevertheless, as discussed later, the results for the {\bf M4} molecule are useful for an understanding of the relationship between ligand fields and structure. Since the magnetic properties were experimentally characterized for the $S_4$-symmetric Cr(IV)({\it o}-tolyl)$_4$ molecule ({\bf M1}), the {\bf M5} molecule with $C_1$ symmetry is {\it not} relevant to our comparison to the experimental data. Therefore, we focus on {\bf M1}, {\bf M2}, and {\bf M3} except for a study of the structural effect on ligand fields.

\section{Methods}

All computations are carried out at the single molecule level. Except for the calculations of the ZPL energies, experimental geometries from Ref.~\citenum{Bayliss2020} are used. For the {\bf M1}, {\bf M2}, {\bf M3}, and {\bf M5} molecules, we use atomic coordinates determined from the x-ray measurements on the corresponding molecular crystals.\cite{Bayliss2020} For the {\bf M4} molecule, the atomic positions obtained from measurements on the diluted molecular crystal formed by diluting Cr(IV)(o-tolyl)$_4$ molecules in their isostructural Sn analogs\cite{Bayliss2020} are used. For all experimental structures, we adjust the C-H bond lengths to 1.09~\AA. For the calculations of the ZPL energies, we relax the experimental geometries of the {\bf M1}, {\bf M2}, and {\bf M3} molecules (in a gas phase) with fixed spins $S=1$ and $S=0$ separately, using the all-electron DFT code {\tt NRLMOL}\cite{NRLMOL,NRLMOL_2} with Gaussian-orbital basis sets and very dense variational mesh under the Perdew-Burke-Ernzerhof (PBE) generalized gradient approximation (GGA)\cite{Perdew1996} for the exchange-correlation functional. The structures are relaxed without any symmetry constraints until the atomic forces are equal to or less than 0.005~eV/\AA.~The atomic coordinates of the relaxed geometries for {\bf M1}, {\bf M2}, and {\bf M3} are listed in Table~S1-S3 in the SI. The DFT-relaxed $S=1$ and $S=0$ geometries are used for the subsequent multireference calculations of the electronic excitations (see below).

The multireference {\it ab-initio} calculations are performed without enforcing any symmetry (i.e., $C_1$ symmetry) using the {\tt Molcas/OpenMolcas} codes.\cite{Molcas,Openmolcas} Scalar relativistic effects are included based on the second-order Douglas-Kroll-Hess Hamiltonian\cite{Douglass1974,Hess1986} and relativistically contracted all-electron correlation-consistent (cc) basis sets.\cite{Dunning1989,deJong2001} For the {\bf M1} and {\bf M4} molecules, we use polarized triple-$\zeta$ (cc-pVTZ-DK) contraction for the Cr atom and all C atoms, while we use polarized double-$\zeta$ (cc-pVDZ-DK) contraction for the H atoms. For the larger {\bf M2} and {\bf M3} molecules, the same basis sets are used except the C atoms from the methyl groups for which we use the cc-pVDZ-DK basis set. We confirm that the slightly smaller basis sets for {\bf M2} and {\bf M3} do not affect our results.

The electronic structure is calculated in three steps. First, in the absence of SOC, both for spin-triplet and spin-singlet states, the spin-free energies and eigenstates are obtained using the state-averaged complete active space self-consistent field (SA-CASSCF) method~\cite{Roos1980,Siegbahn1981}. In the second step, the spin-free energies are further improved by including dynamic correlations, using the multi-state~\cite{Finley1998} second-order multireference perturbation theory (CASPT2)~\cite{Andersson1990,Andersson1992}. These CASPT2-corrected spin-free energies are shown as electronic VE energies in Fig.~\ref{fig:spectrum}(a). By applying the same multireference {\it ab-initio} methods to the DFT-relaxed geometries, we obtain the ZPL energies (see Table~\ref{tab:2}). Finally, in the third step, the SOC is included within the atomic mean-field approximation~\cite{Hess1996} using the restricted active space state interaction (RASSI)~\cite{Malmqvist2002} method.

\begin{figure}[htb!]
\centering
\includegraphics[width=0.49\linewidth]{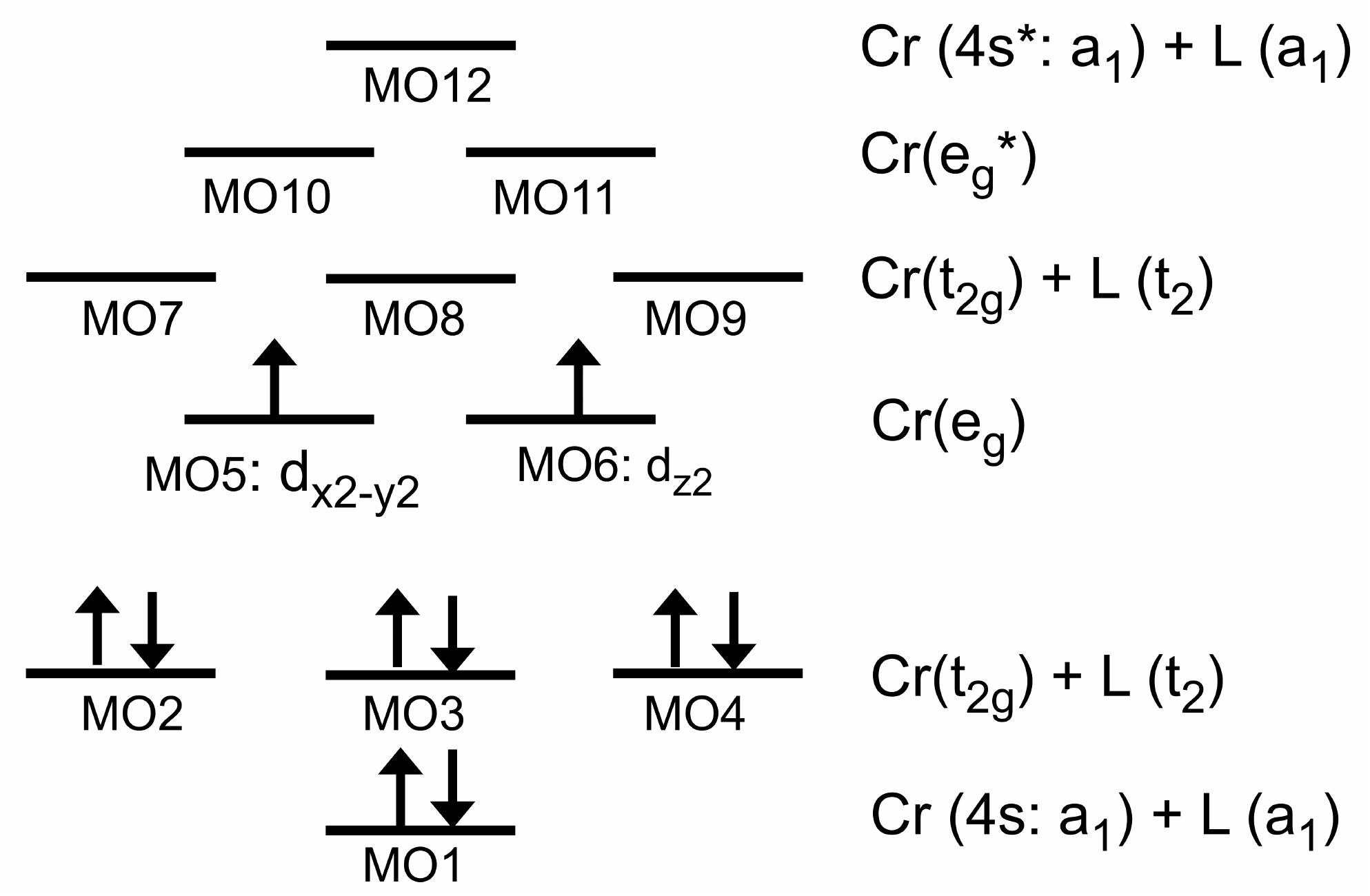}
\caption{List of active molecular orbitals (MOs) used for the CASSCF(10,12) calculations for all considered molecules, where ''L'' stands for ligands. The configuration on the left hand side shows the ground spin-triplet state with nominal orbital occupancy. The active orbital images for the ground spin-triplet state for {\bf M1} are shown in Fig.~S1 in the SI.}
\label{fig:CAS}
\end{figure}

In order to determine an optimal active space for the SA-CASSCF calculations, we start with a simple picture of Cr$^{4+}$ ion with 2 active electrons in 5 active 3$d$ orbitals. We use Cartesian axes along the axes of the cube in which the quasi-tetrahedral carbon cage (Fig.~\ref{fig:geo}) is inscribed. In the tetrahedral symmetry, the $d_{xy}$, $d_{xz}$, and $d_{yz}$ orbitals (MO7, MO8, MO9) strongly hybridize with three ligand orbitals (MO2, MO3, MO4). Therefore, these ligand orbitals and the corresponding 6 electrons (the ligand orbitals are nominally doubly occupied) are included in the active space. The $d_{x^2-y^2}$ and $d_{z^2}$ orbitals (MO5, MO6) are non-bonding and they largely retain their atomic-like character. For a proper description of strong radial correlations for such localized orbitals, we add two (nominally empty) double-$d$-shell orbitals of the same spherical symmetry to the active space: $d^*_{x^2-y^2}$ and $d^*_{z^2}$ (MO10, MO11). Finally, the active space is supplemented by nominally empty Cr 4$s$-like orbital (MO12) and the nominally doubly occupied ligand orbital (MO1) that hybridizes with the Cr 4$s$-like orbital. Therefore, the active space consists of 10 electrons in 12 orbitals and we denote it as CAS(10,12). Figure~\ref{fig:CAS} summarizes the active molecular orbitals and depicts the configuration of the ground spin-triplet state with nominal orbital occupancy. The active orbital images for the spin-triplet state of the {\bf M1} molecule are shown in Fig.~S1 in the SI. 

The number of roots in the SA-CASSCF(10,12) calculations is determined based on a significant energy gap between root energies as well as whether the resulting orbitals and their natural occupations (or energies) reflect the symmetry of the considered molecule. Indeed, only with the orbitals that retain the molecular symmetries for both the $S=1$ and $S=0$ states, we can accurately determine the ZFS $D$ and $E$ parameters. For instance, for the {\bf M1} molecule with $S_4$ symmetry, the orbitals belonging to the same $E$ irreducible representation must have equal state-averaged natural occupancies (active orbitals) or energies (inactive orbitals), and consequently, the calculated $E$ parameter must be zero. Therefore, we use 13 (15) roots for the state averaging in SA-CASSCF(10,12) calculations for the $S=1$ ($S=0$) case. The dominant configurations of the 13 spin-triplet roots and 15 spin-singlet roots for {\bf M1} are listed in Table~S8-S9 in the SI, respectively. For the $S=1$ ($S=0$) case, there is a significant energy gap between the 13$^\text{th}$ and 14$^\text{th}$ (15$^\text{th}$ and 16$^\text{th}$) roots. More importantly, for both spin states, molecular orbitals preserve the molecular symmetries for all the considered molecular structures (experimental and DFT-relaxed geometries). In particular, for the {\bf M1} molecule, the calculated $E$ parameter is zero as required by the $S_4$ symmetry. For {\bf M1} molecule, we also check that the ZFS $D$ and $E$ parameters do not change much compared to the values in Table 3, when for the S=1 state, {\it 14 roots} are used in the state average procedure instead.

The multi-state CASPT2 calculations are done for all 13 (15) roots for the spin-triplet (spin-singlet) state using the default ionization potential-electron affinity (IPEA) shift (0.25 a.u.)~\cite{Ghigo2004}. In order to improve the convergence, an additional real shift~\cite{Roos1995} of 0.3 a.u. is used in the CASPT2 calculations for all considered molecules. We check that our results are insensitive to the shift value.

The RASSI-SOC calculations are performed within the space spanned by 13 spin-triplet roots (including their spin sub-levels) and 15 spin-singlet roots using the SA-CASSCF wavefunctions and the multi-state CASPT2-corrected energies. The ZFS (or magnetic anisotropy) effective spin Hamiltonian as well as the ${\bf g}$ tensor are then constructed using the SINGLE$\_$ANISO approach~\cite{Chibotaru2012}.

The hyperfine coupling and superhyperfine coupling parameters are calculated using the method described in Ref.~\citenum{Wysocki2020}. Since we focus on the coupling parameters of the ground state, they are evaluated based on CASSCF(10,12) calculations for a single (ground spin-triplet) root. Correspondingly, RASSI-SOC calculations are done within the space spanned only by the three spin sub-levels of the ground spin-triplet using both CASSCF wavefunctions and energies. In this case, the CASPT2 step is skipped because it only affects the energy spacing between the roots.

The contribution of dipolar electron spin-spin interaction to the ZFS parameters for the {\bf M1} molecule is calculated using the {\tt ORCA} code~\cite{Orca}. We use cc-pVDZ-DK basis set for all atoms and CASSCF(8,8) calculations with MO2-MO9 as active orbitals for the ground spin-triplet state.

\section{Results \& Discussion}

{\bf \large Multireference Electronic Excitations}

In order to include static and dynamic correlations into the electronic structure, we perform multireference {\it ab-initio} calculations, using the state-average complete active space self consistent field (SA-CASSCF) method \cite{Roos1980,Siegbahn1981} followed by the multi-state\cite{Finley1998} second-order multireference perturbation theory correction (CASPT2),\cite{Andersson1990,Andersson1992} as implemented in the {\tt Molcas/Openmolcas} codes.\cite{Molcas,Openmolcas} The active space consists of 10 electrons and 12 molecular orbitals. A computational detail is described in Methods section.

\begin{figure}[H]
\centering
\includegraphics[width=1.0\linewidth]{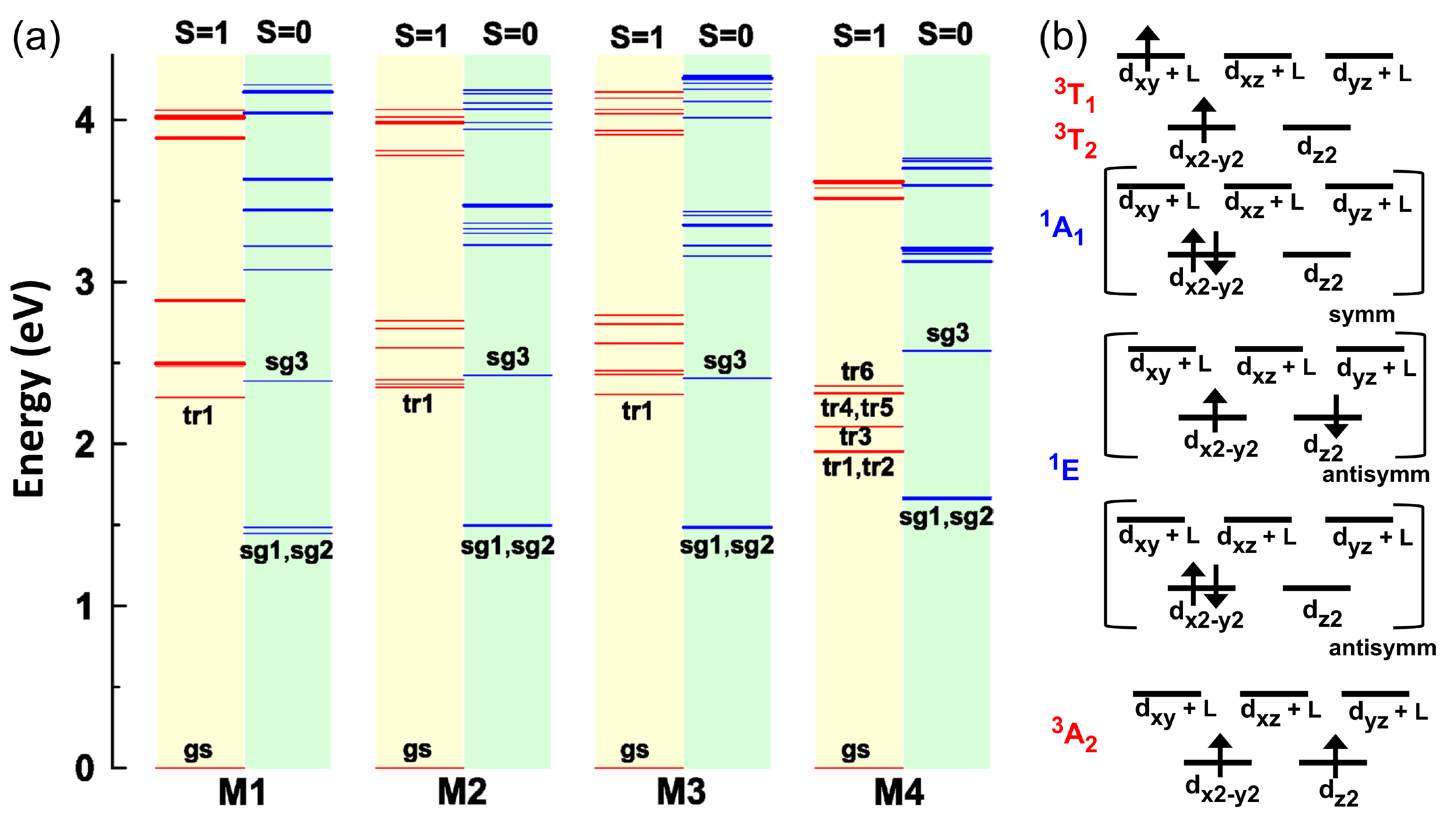}
%\hspace{1.0truecm}
\includegraphics[width=0.9\linewidth]{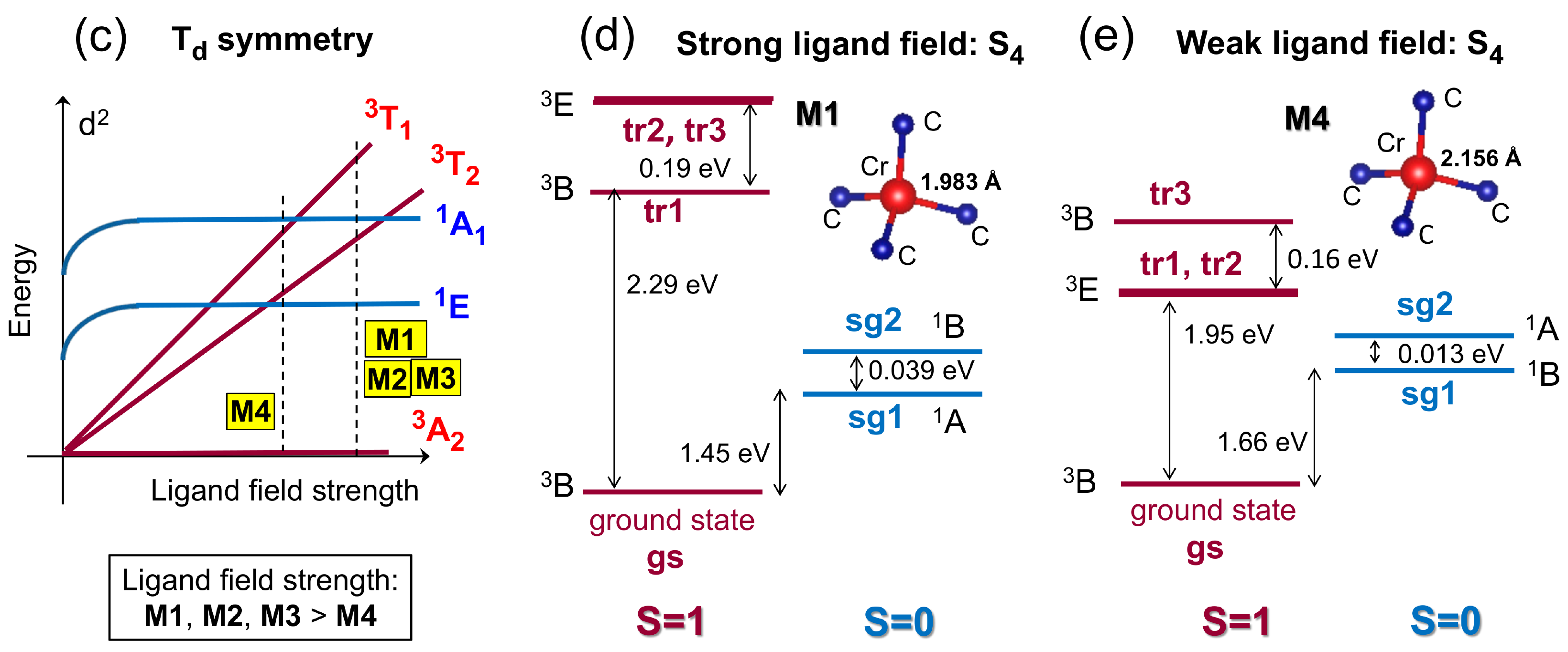}
\caption{(a) Multireference electronic excitation energies of all considered molecules with the experimental structures without SOC. The ground state and first-excited $S=1$ state are labeled as {\bf gs} and {\bf tr1}, respectively, while three lowest $S=0$ states are labeled as {\bf sg1}, {\bf sg2} and {\bf sg3}. Thicker lines indicate states with degeneracy. (b) Diagrams of low-lying state configurations under $T_d$ symmetry, considering only Cr 3$d$ orbitals, for simplicity (L: ligand orbitals), where {\bf antisymm} and {\bf symm} denote antisymmetric and symmetric combinations of two possible configurations. For example, one of the $^1E$ states is an antisymmetric combination of two electrons at either $d_{x^2-y^2}$ or $d_{z^2}$ orbital. For $^3T_1$ and $^3T_2$, six configurations are possible from the diagrams. (c) Schematic Tanabe-Sugano diagram \cite{Tanabe1954} of energy vs ligand-field strength for 3$d^2$ systems under $T_d$ symmetry. The lower molecular symmetries than $T_d$ lift threefold degeneracy in $^3T_1$ and $^3T_2$ and twofold degeneracy in $^1E$. For {\bf M4}, {\bf tr1}-{\bf tr6} levels appear between {\bf sg3} and {\bf sg1}/{\bf sg2} levels in energy. That is not the case for the others. (d)-(e) four spin-triplet and two spin-singlet states for {\bf M1} and {\bf M4}. The ordering of the excited spin-triplet (spin-singlet) states for {\bf M4} is reversed to that for {\bf M1}.}
\label{fig:spectrum}
\end{figure}

Figure~\ref{fig:spectrum}(a) shows our calculated SA-CASSCF+CASPT2 energies of spin-triplet and spin-singlet states without SOC for all considered molecules with the experimental atomic structures taken from Ref.~\citenum{Bayliss2020}. In all cases, we observe the following features: (i) the ground state, {\bf gs}, is a spin-triplet ($S=1$) well separated in energy from the first-excited spin-triplet state, {\bf tr1}; (ii) the first excited state, {\bf sg1}, is a spin-singlet ($S=0$) almost degenerate with the next spin-singlet state, {\bf sg2}. Our calculated low-lying spin-triplet and spin-singlet energies are listed in Table~S4 of the SI.

Figure~\ref{fig:spectrum}(b) presents diagrams of a few low-lying state configurations under exact $T_d$ symmetry, for simplicity, considering only Cr 3$d$ orbitals, where {\bf L} stands for ligand 2$p$ orbitals hybridized with Cr $t_{2g}$ orbitals. Under exact $T_d$ symmetry, the ground state, $^3A_2$, consists of nominally singly-occupied lower-energy $e_g$ orbitals and nominally empty higher-energy $t_{2g}$ orbitals hybridized with ligand $p$ orbitals. For strong ligand fields, the ligand field theory\cite{Tanabe1954} dictates that the first- and second-excited states are spin-singlet states with $^1E$ and $^1A_1$ characters, respectively (Fig.~\ref{fig:spectrum}(b),(c)). For {\bf M1}, since the symmetry is lowered to $S_4$, the ground state has now a character of $^3B$, while the degenerate lowest spin-singlet states are split into $^1A$ and $^1B$ states (Fig.~\ref{fig:spectrum}(d)). The bottom and top diagrams of the $^1E$ state shown in Fig.~\ref{fig:spectrum}(b) correspond to the $^1A$ and $^1B$ states, respectively. The $^3T_2$ state under $T_d$ symmetry is also split into $^3B$ and $^3E$ characters under $S_4$ symmetry. Interestingly, for {\bf M4} (with $S_4$ symmetry), all six lowest excited triplet states, {\bf tr1}-{\bf tr6}, that correspond to two threefold degenerate states ($^3T_2$, $^3T_1$) under $T_d$ symmetry, appear between the excited spin-singlet state {\bf sg3} and the almost degenerate singlet states {\bf sg1}/{\bf sg2} in energy. However, that is not the case for the other molecules (Fig.~\ref{fig:spectrum}(a)). The Tanabe-Sugano diagram (Fig.~\ref{fig:spectrum}(c)) indicates that {\bf M4} has a weaker ligand field than the rest of the molecules, which is consistent with the Cr-C bond length being significantly larger for {\bf M4}. We also find that different ligand-field strengths affect the ordering of the excited spin-triplet states ($^3B$ and $^3E$) as well as that of the excited spin-singlet states ($^1A$ and $^1B$) for {\bf M1} and {\bf M4}. Both orderings are reversed in the two molecules (Fig.~\ref{fig:spectrum}(d),(e)).

We extract a vertical excitation (VE) energy for each molecule from an energy difference between the ground state and the lowest spin-singlet state (Fig.~\ref{fig:spectrum}(a)) obtained from the experimental geometry. Note that even though the transition is between the states with different spin multiplicities, we still use the term VE since energies of both states are calculated for the same atomic structure. Such definition of the VE energy is consistent with that used for point defects in large-gap semiconductors.\cite{Bockstedte2018,Ma2020,Sajid2020}. Our calculations show that the VE energies for {\bf M1}, {\bf M2}, and {\bf M3} are close to each other, in the range of 1.448-1.493~eV, while the VE energy of {\bf M4} is 1.658~eV (Table~\ref{tab:2}). In order to understand what causes larger VE energy for the \textbf{M4} molecule, we perform SA-CASSCF+CASPT2 calculations for a slightly modified {\bf M4} structure (referred to as {\bf M4'}) where the Cr-C bond length is set to the same as that of {\bf M1}, 1.983~\AA,~while the structures of the individual aryl ligands remain unchanged. In other words, the {\bf M4'} structure retains S$_4$ symmetry and the aryl ligands in it are simply brought closer to the Cr site. We find that the VE energy for {\bf M4'} is 1.459~eV, which is close to that of {\bf M1}. This indicates that the higher VE energy for the \textbf{M4} molecule is due to a larger Cr-C bond length. Indeed, a larger Cr-C bond length results in a weaker hybridization of the Cr $3d$ orbitals with the ligand orbitals. Consequently, the $3d$ orbitals are more localized at the Cr site and the electron-electron repulsion among the $3d$ electrons increases. This leads to a larger intra-site exchange interaction and higher triplet-singlet VE energy. This mechanism is confirmed by calculating the Racah parameters that describe the inter-electronic repulsion within the Cr $3d$ shell (see Table~S16 in SI). We find that, indeed, the \textbf{M4} molecule has larger Racah parameters than the \textbf{M1} molecule.

\begin{table}[htb!]
\centering
\caption{Calculated vertical excitation (VE) energies $\Delta E_{\text{VE}}$ between the ground spin-triplet state ({\bf gs}) and the lowest spin-singlet state ({\bf sg1} in Fig.~\ref{fig:spectrum}) as well as ZPL energies $\Delta E_{\text{ZPL}}$ of all considered molecules (see the main text for detail).}
\begin{tabular}{c|c|c|c|c}
\hline \hline
Properties                          & \textbf{M1}  & \textbf{M2}    & \textbf{M3}    & \textbf{M4} \\
\hline
$\Delta E_{\text{VE}}$  (eV)        & 1.448        &  1.493         &  1.483         & 1.658 \\
$\Delta E_{\text{ZPL}}$ (eV)        & 1.415        &  1.436         &  1.419         & N/A \\
$\Delta E_{\text{ZPL}}$ (eV) (Exp.~\citenum{Bayliss2020})  & 1.210        &  1.229         &  1.210         & N/A \\ \hline
\hline
\end{tabular}
\label{tab:2}
\end{table}

We also compute a ZPL energy by subtracting a SA-CASSCF+CASPT2 energy of the ground spin-triplet state using a DFT-relaxed geometry with $S=1$ from that of the lowest spin-singlet state using a separate DFT-relaxed geometry with $S=0$ (see Methods section for detail). We confirm that the DFT-relaxed structures of {\bf M1}, {\bf M2}, and {\bf M3} in the $S=1$ state are very close to the experimental structures of {\bf M1}, {\bf M2}, and {\bf M3}. Quantitative comparison of the structures is listed in Table S4 in the SI. We also check that the SA-CASSCF+CASPT2 energies involving the DFT geometries are very similar to those from the experimental geometries for the Cr-based molecular crystals (see Table S7 in the SI). Upon the geometry relaxation, we observe large structural changes only in the {\bf M4} molecule. As a result, the relaxed atomic coordinates of {\bf M4} become essentially identical to those of {\bf M1} (see Table S5 in the SI). Henceforth, we investigate properties of {\bf M1}, {\bf M2}, and {\bf M3} only. We find that the ZPL energy of each molecule is smaller by $< 0.1$~eV than the VE energy (Table~\ref{tab:2}). The calculated ZPL energies of the considered molecules are slightly higher than the experimental ZPL energies which are in the range of 1.210-1.229~eV.\cite{Bayliss2020} These small deviations from the experimental energies are typical for SA-CASSCF+CASPT2 calculations due to approximate treatment of dynamical correlations\cite{ZLi2008,Schapiro2013} as well as inaccuracy in the DFT-relaxed singlet structures. As a point of interest, we also report that our DFT calculations show that the $S=1$ state always has a lower energy than the $S=0$ state, and that the DFT-calculated ZPL values, i.e., the DFT energy differences between the DFT-relaxed singlet and triplet structures, are in the range of 1.07-1.11 eV, for the {\bf M1}, {\bf M2}, and {\bf M3} molecules (see Table S8 in the SI).

\begin{figure}[htb]
\centering
\includegraphics[width=0.5\linewidth]{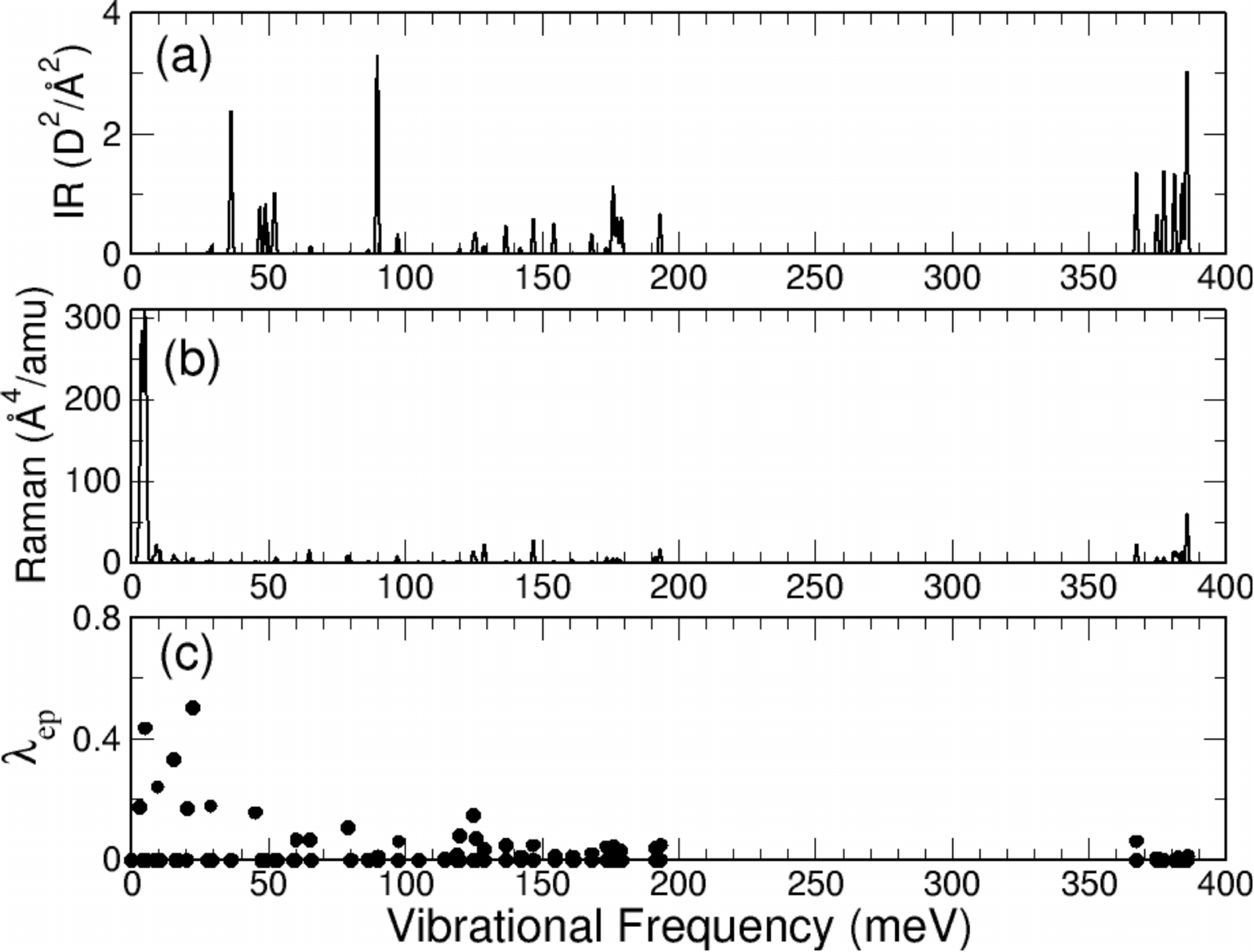}
\caption{(a) Infrared intensity (in units of (Debye/\AA)$^2$) and (b) Raman scattering activity (in units of \AA$^4$/amu) of the vibrational modes for {\bf M1} which are computed using DFT. (c) Calculated dimensionless vibronic coupling strength of the vibrational modes for the ground spin-triplet state of {\bf M1}.}
\label{fig:vib}
\end{figure}

Since the difference between the VE and ZPL energies is small (Table~\ref{tab:2}), it is likely that electron-phonon coupling (i.e., vibronic coupling) may be small for the ground spin-triplet state and lowest spin-singlet state of the {\bf M1}, {\bf M2}, and {\bf M3} molecules. As a representative, we compute vibrational modes and vibronic coupling for the ground spin-triplet state of {\bf M1}, using the DFT code {\tt NRLMOL}.\cite{NRLMOL,NRLMOL_2,Briley1998} As shown in Fig.~\ref{fig:vib}, the vibrational modes have a range of 2.8~meV to 386 meV. We also calculate dimensionless vibronic coupling $\lambda_{\text{ep}}$ of each vibrational mode using the method described in Ref.~\citenum{McCaskey2015}. We find that only 10 vibrational modes (2.8-124.8~meV) with $A$ symmetry have intermediate dimensionless vibronic coupling strength ($0.1 < \lambda_{\text{ep}} < 0.5$), while the rest of the modes have much smaller or zero vibronic coupling. This result is consistent with our result that the VE and ZPL energies differ by less than 0.1~eV. The aforementioned structural and energetic agreements between the DFT relaxed and experimental geometries, suggest that the DFT-calculated vibrational spectra and vibronic couplings should be reasonably accurate. So far, vibrational spectra (infrared and Raman spectra) of the considered molecules have not been experimentally measured.
\\
\\

{\bf \large Spin-Orbit Coupling Effect I: Zero-Field Splitting}

\begin{table}[t!]
\centering
\caption{Calculated zero-field splitting or magnetic anisotropy parameters $D$ and $E$ (in GHz) as well as {\bf g} tensor of the ground spin-triplet state for all considered molecules with the experimental geometries, compared to experimental data from Ref.~\citenum{Bayliss2020}. The experimental ${\bf g}$ tensor is isotropic and it is 1.985 for the \textbf{M1}, \textbf{M2}, and \textbf{M3} molecules. The theoretical $D$, $E$, and ${\bf g}$ tensor are from the RASSI-SO-CASPT2 calculations with 13 spin-triplet and 15 spin-singlet states.}
\begin{tabular}{c|r|r|r}
\hline \hline
Properties                       & \textbf{M1}  & \textbf{M2}   & \textbf{M3}   \\
\hline
$D$                              & $-$5.7       & $-$3.3        & $-$2.4        \\
$E$                              & 0.0          & 0.7           & 0.3           \\
$|E/D|$                          & 0.0          & 0.21          & 0.13          \\ \hline
$|D|$  (Exp.~\citenum{Bayliss2020}) & 3.63         & 1.83          & 4.11       \\
$|E|$  (Exp.~\citenum{Bayliss2020}) & 0.00         & 0.49          & 0.54       \\
$|E/D|$                          & 0.0          & 0.27          & 0.13          \\ \hline
$g_{xx}$                         & 1.983        & 1.979         & 1.983         \\
$g_{yy}$                         & 1.983        & 1.980         & 1.981         \\
$g_{zz}$                         & 1.977        & 1.982         & 1.978         \\
\hline \hline
\end{tabular}
\label{tab:3}
\end{table}

All the electronic spin-triplet states shown in Fig.~\ref{fig:spectrum} are split by SOC and/or dipolar electron spin-spin coupling (SSC). In this work, we mostly focus on the level splitting of the ground spin-triplet state (non-degenerate) which can be described by the following ZFS Hamiltonian:
\begin{equation}
\hat{H}_\text{eff}=D\hat{S}_z^2-\frac{1}{3}DS(S+1)+E\big(\hat{S}_x^2-\hat{S}_y^2\big).
\label{eq:ZFS}
\end{equation}
Here, $D$ and $E$ are uniaxial and transverse (rhombic) ZFS (or magnetic anisotropy) parameters, and $\hat{\mathbf{S}}$ is the pseudospin operator with $S=1$. The coordinate system corresponds to the molecular anisotropy axes. For $S=1$, the eigenvalues of the ZFS Hamiltonian are $D+E$, $D-E$, and zero. 

We first investigate a SSC contribution to the ZFS parameters of the ground spin-triplet state for the {\bf M1} molecule with the experimental geometry, by performing multireference calculations using the {\tt ORCA} code~\cite{Orca}. The SSC contribution turns out to be negligible ($|D_{\text{SSC}}|<$0.01cm$^{-1}$, $E_{\text{SSC}}=0$). As expected, since the SSC contribution is a first-order effect, it was shown that the contribution was mainly determined by the ground state rather than excited states, and that it was not sensitive to the active space size, the number of roots for state average, and inclusion of dynamic correlations.\cite{Bhandari2021} Therefore, we expect that the SSC contributions for the other molecules are also negligible and that the experimental ZFS parameters originate entirely from SOC.

Since the ground spin-triplet state is orbitally non-degenerate, it is split by second-order SOC, which depends on the excited states. We compute SOC contributions to the $D$ and $E$ parameters of the ground spin-triplet state for all considered molecules with the experimental atomic coordinates, considering 13 spin-triplet and 15 spin-singlet states for each molecule (see Methods section for detail). As listed in Table~\ref{tab:3}, for the {\bf M1} molecule, the calculated $E$ parameter is zero, which reflects $S_4$ symmetry. On the other hand, for the {\bf M5} molecule, the calculated $E$ parameter is non-zero, 0.8 GHz, due to its lower symmetry, $C_1$. The calculated $|D|$ parameter for {\bf M5} is slightly larger than that for {\bf M1} (See Table~S6 in the SI). The magnitude of the calculated $D$ parameter of {\bf M1} is comparable to the experimental value,\cite{Bayliss2020} while those of {\bf M2} and {\bf M3} are somewhat overestimated and underestimated than the experimental values,\cite{Bayliss2020} respectively. Note that the $|D|$ parameter value of the considered molecules is 1-2 orders of magnitude smaller than that of typical 3$d$ transition-metal complexes\cite{Duboc2010,Bucinsky2019} and that the deviations of our calculated $|D|$ values from the experimental ones are smaller than those in Refs.~\citenum{Duboc2010,Bucinsky2019}. Interestingly, the calculated $|E/D|$ ratios for {\bf M1}, {\bf M2}, and {\bf M3} agree well with the experimental values,\cite{Bayliss2020} which suggests that the active spaces retain the molecular symmetries. The small discrepancies between the calculated and the experimental ZFS parameters may be attributed to (i) atomic structures that could slightly differ from experimental systems and (ii) approximate treatment of dynamical correlations which influences the excited-state wavefunctions and energies. In particular, correlations involving $\pi$ ligand orbitals may play an important role. Since the calculated $D$ and $E$ parameters in Table~\ref{tab:3} arise from the second-order SOC effect, they depend on high-energy spin-singlet states as well as spin-triplet states. Importantly, SOC contributions to $D$ and $E$ from different excited states have different magnitudes and different signs (see Tables~S5-S7 in the SI). (The configurations of all roots for {\bf M1} are listed in Tables~S8-S9 in the SI.) As a result, the values of the ZFS parameters are, to a large degree, determined by high-energy multireference excited states that can be substantially modified by slight changes of ligands. The ZFS parameter values are, therefore, expected to be sensitive to small perturbations or atomic structure variations. This is consistent with experimental and theoretical findings that the $D$ and $E$ parameters noticeably vary among the \textbf{M1}, \textbf{M2}, and \textbf{M3} molecules.

Furthermore, we find that the signs of the calculated $D$ parameters are {\it negative} for all considered molecules. This indicates that the lowest magnetic sub-levels are degenerate $M_s=\pm$1 states for $E=0$, or linear combinations of $M_s=\pm$1 states for $E\neq0$. While the sign of the $D$ parameter was {\it not} determined in the experimental studies,\cite{Bayliss2020,Laorenza2021} it can be unambiguously obtained by specific heat measurements as a function of temperature\cite{Rubin2021} as well as electron paramagnetic resonance experiments under high external magnetic fields ($\sim$several tesla).\cite{Krzystek2006}

For the {\bf M1} molecule, the magnetic easy axis is found to be along the $S_4$ symmetry axis. Anisotropy in the calculated ${\bf g}$ tensor (Table~\ref{tab:3}) corroborates the magnetic anisotropy or non-zero ZFS parameters. The principal values of the calculated ${\bf g}$ tensor agree with the experimental values, although the latter values were assumed to be isotropic in fitting of the experimental electron paramagnetic resonance spectra.\cite{Bayliss2020} The magnetic axes which diagonalize the ${\bf g}$ tensor, in general, do not coincide with the magnetic anisotropy axes.\cite{Chibotaru2012} As a result, the anisotropy in the ${\bf g}$ tensor in Table~\ref{tab:3} does not need to be strictly correlated with the sign of the ZFS $D$ parameter, similarly to Refs.~\citenum{Limburg2001,Mossin2001,Tregenna2003}.

We also briefly discuss ZFS of the first excited non-degenerate spin-triplet state since its spin sub-levels $M^{\prime}_s$ are involved in ISC, as discussed in the next section. Regarding the first excited spin-triple state, we consider only the second-order SOC contribution to its ZFS parameters which are labeled as $D^{\prime}$ and $E^{\prime}$ in order to distinguish from those of the ground state. Following the same procedure as above, we find that for the {\bf M1} and {\bf M3} molecules, the $D^{\prime}$ parameter is negative, while for the {\bf M2} molecule, it is positive. Similarly to the case of the ground spin-triplet state, the {\bf M2} and {\bf M3} molecules have a nonzero $E^{\prime}$ parameter, while it vanishes for the {\bf M1} molecule. Compared to the ground spin-triplet state, the splitting of the sub-levels is much larger. For the three molecules, the $|D^{\prime}|$ value ranges from 72.8 to 91.7~GHz (2.43-3.06~cm$^{-1}$), while the nonzero $E^{\prime}$ value ranges from 5.1 to 20.7~GHz (0.17-0.69~cm$^{-1}$). 
\\

{\bf \large Spin-Orbit Coupling Effect II: Intersystem Crossing}

\begin{figure}[htb!]
\centering
\includegraphics[width=1.0\linewidth]{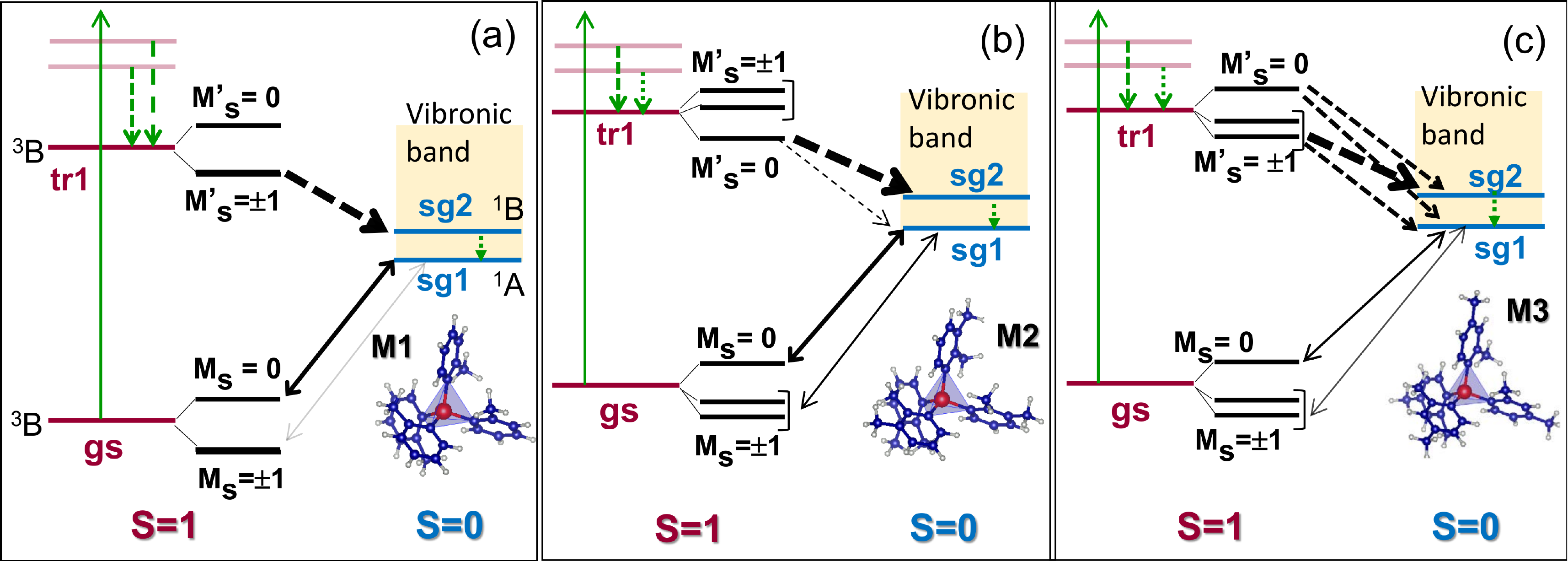}
\caption{Diagrams of radiative (solid arrows) and non-radiative (dashed arrows) relaxation/absorption processes for (a) {\bf M1}, (b) {\bf M2}, and (c) {\bf M3}. Spin-conserving (spin-flip) processes are shown as green (black) arrows. Thicknesses of black dashed arrows represent magnitudes of the SOC matrix elements, while thicknesses of black solid arrows denotes magnitudes of the oscillator strengths. The green solid up-arrow indicates off-resonant excitation. Due to fast internal conversion from {\bf sg2} and {\bf sg1}, radiative transitions between {\bf sg2} and the $M_s$ sub-levels of {\bf gs} are not shown. Shaded areas in the spin-singlet states, {\bf sg1} and {\bf sg2}, represent vibronic bands. In the case of off-resonant excitation, the predominant radiative relaxation from {\bf sg1} to the $M_s=0$ sub-level of {\bf gs} renders the spin initialization into the $M_s=0$ sub-level of {\bf gs}, for all considered molecules.}
\label{fig:ISC}
\end{figure}

In addition to the level splitting, SOC allows for radiative and non-radiative transitions between spin-triplet and spin-singlet states. Such transitions plays an important role in achieving an optical spin interface. In particular, in Ref.~\citenum{Bayliss2020} it was proposed that an optical spin initialization for the considered Cr(IV)(aryl)$_4$ molecules can be realized by applying a resonant light with frequency equal to the energy difference between the lowest singlet state \textbf{sg1} and one of the sub-levels of the spin-triplet ground state. The molecule in this ('bright') spin sub-level is then more likely to be excited into \textbf{sg1} than if it is in other ('dark') spin sub-levels (assuming that the line broadening is smaller than the sub-level separation). On the other hand, the molecule in \textbf{sg1} can decay into any of the ground triplet spin sub-levels with comparable rates, producing the photoluminescence. Since the lifetime of \textbf{sg1} is much smaller than the spin-lattice relaxation time for the ground triplet state,\cite{Bayliss2020} during optical cycles, these processes would eventually transfer the molecule from the 'bright' spin sub-level to the 'dark' spin sub-level(s). Such depopulation of the 'bright' spin sub-level was demonstrated experimentally for the \textbf{M1} molecule that was reflected in a 14\% reduction of the observed photoluminescence.\cite{Bayliss2020}

Our calculations provide additional insight into this process by evaluating absorption and emission rates between the spin-singlet states and the ground spin-triplet sub-levels $M_s$. We find that the decay rates from {\bf sg1} to the different $M_s$ sub-levels of {\bf gs} are not always the same. In particular, for the $S_4$-symmetric {\bf M1} molecule, a radiative transition occurs predominantly between the $M_s=0$ sub-level and {\bf sg1} (denoted as a thick solid black arrow in Fig.~\ref{fig:ISC}(a)), while the $M_s=\pm1$ levels have radiative transitions primarily with {\bf sg2} (not shown). If the {\bf M1} molecule is excited from the lowest spin-triplet sub-levels $M_s=\pm1$ to {\bf sg2} by a light with a resonant frequency, then instead of a radiative decay back to the $M_s=\pm1$ sub-levels, the molecule undergoes non-radiative internal conversion from {\bf sg2} to {\bf sg1} (denoted as a tiny dashed green arrow in Fig.~\ref{fig:ISC}(a)) and a subsequent radiative decay from {\bf sg1} to the $M_s=0$ sub-level, since the internal conversion rate is much larger than the radiative decay rate. During optical cycles, the $M_s=\pm1$ sub-levels become depopulated and the molecule is initialized to have $M_s=0$. The same mechanism is applied to the {\bf M2} and {\bf M3} molecules (Fig.~\ref{fig:ISC}(b),(c)). As the molecular symmetry is lowered, the ratio between the radiative decay to the $M_s=0$ sub-level and that to the $M_s=\pm1$ sub-levels greatly decreases, and therefore the efficiency in the optically-induced spin initialization decreases. Based on this, we expect that the efficiency in the optical spin initialization is largest for {\bf M1} and smallest for {\bf M3}.

For point defects in wide bandgap semiconductors, an optical spin interface is realized using a different approach that utilizes non-radiative ISC transitions.\cite{Goldman2015,Soykal2016,Doherty2013,Galireview2019} Our results indicate that a similar technique may be used to achieve optical spin initialization in the Cr(IV)(aryl)$_4$ molecules which would be an alternative to the method used in Ref.~\citenum{Bayliss2020}. First, a non-resonant light is used to excite the molecules into higher-lying spin-triplet states (see full green arrows in Fig.~\ref{fig:ISC}), which is followed by rapid non-radiative internal conversion to {\bf tr1} (dashed green arrows in Fig.~\ref{fig:ISC}). In principle, spontaneous emissions from {\bf tr1} to {\bf gs} can then occur. Our estimation, however, shows that a spontaneous emission rate from a higher-energy spin-triplet state to {\bf gs} is at most $\sim$10$^6$~s$^{-1}$. Since this rate is much smaller than typical non-radiative ISC rates in TM-based molecules, $\sim$10$^{12}$~s$^{-1}$,\cite{Fataftah2020,Juban2005,Dorn2020} the radiative decay from the higher-lying spin-triplet states to {\bf gs} is unlikely to be observed. Instead, the molecules undergo non-radiative ISC from the $M^{\prime}_s$ sub-levels of {\bf tr1} to either {\bf sg1} or {\bf sg2}. This ISC transition involves phonon emission when the energy of {\bf tr1} falls within the phonon or vibronic energy range of {\bf sg1}/{\bf sg2} (shaded areas in Fig.~\ref{fig:ISC}). A full quantitative calculation of the ISC rates requires consideration of the direct SOC term and two complex terms arising from spin-vibrational couplings, i.e., Eqs.(18)-(20), in Ref.~\citenum{Penfold2018}. Such calculations are beyond the scope of the current work. However, in order to figure out dominant ISC channels, computation of the SOC matrix elements is often sufficient.\cite{Soykal2016,Evans2018} We provide the SOC matrix elements, $|\langle \Psi_f|{\cal H}_{\rm SOC}|\Psi_i\rangle|$, between the {\bf tr1} $M_s'$ level and {\bf sg2} (or {\bf sg1}) for the {\bf M1}, {\bf M2}, and {\bf M3} molecules in Table S14 in the SI. The relative importance of the non-radiative ISC transition rates is estimated by comparing the SOC matrix elements between the $M^{\prime}_s$ sub-levels of {\bf tr1} and the spin-singlet states. As shown in Fig.~\ref{fig:ISC}, non-radiative ISC transition rates (denoted as black dashed arrows) strongly depend on both characters of the sub-levels $M^{\prime}_s$ of {\bf tr1} and the ligand type. For example, for the {\bf M1} molecule, the ISC transition occurs predominantly between the $M^{\prime}_s=\pm1$ levels of {\bf tr1} and {\bf sg2}. Such a $M^{\prime}_s$-selective ISC rate plays a crucial role in establishing the spin initialization in the NV$^{-}$ center and similar systems\cite{Goldman2015,Soykal2016,Doherty2013,Galireview2019} because the spontaneous emission rate from {\bf tr1} to {\bf gs} is comparable to the ISC rate. In order to realize such a mechanism within molecules, substantial reduction of the ISC rate is required.\cite{Dorn2020,Wegeberg2021} For the considered Cr(IV)(aryl)$_4$ molecules, the $M^{\prime}_s$-selective ISC rate, however, does not affect the spin initialization. Nevertheless, the spin initialization can be potentially achieved here due to the $M_s$-selective phosphorescence discussed above. Indeed, the dominant pathway involves internal conversion to {\bf sg1} and then primarily radiative relaxation to $M_s$=0 of {\bf gs}, which under optical cycles results in transfer of population from $M_s=\pm1$ to $M_s=0$ for all considered molecules. Importantly, the efficiency of this process can be improved by ligand manipulations since the $M_s$-selectivity of phosphorescence highly depends on ligands.
\\

{\bf \large (Super)Hyperfine Coupling: Spin Decoherence}

The molecular electronic spin states are susceptible to various sources of spin decoherence arising from environment\cite{Wolfowicz2021} even at low temperatures. In molecular crystals, electron-phonon (and spin-phonon) interactions are known to be critical for electron spin relaxation time $T_1$ which limits electron spin coherence time $T_2$. The latter time was shown to increase by lowering the concentration of Cr(IV)-based molecules in diluted molecular crystals,\cite{Laorenza2021} suggesting that dipolar electron spin-spin interactions between different Cr(IV)-based molecules play an important role in spin decoherence. At the single molecule level, the $^{53}$Cr nuclear spin and $^{13}$C and $^1$H nuclear spins of the ligands interact with the molecular electronic spin. The interaction between the molecular electronic spin and the nuclear spins is described by the following effective spin Hamiltonian
\begin{equation}
\hat{H}_\text{HF}=\hat{\mathbf{S}}\cdot\mathbf{A}_\text{Cr}\cdot\hat{\mathbf{I}}_\text{Cr} + \hat{\mathbf{I}}_\text{Cr}\cdot\mathbf{P}_\text{Cr}\cdot\hat{\mathbf{I}}_\text{Cr}+
\sum_{i\text{=all $^1$H,$^{13}$C}}\hat{\mathbf{S}}\cdot\mathbf{A}_i\cdot\hat{\mathbf{I}}_i.
\label{eq:SHF}
\end{equation}
The first term is the magnetic hyperfine interaction between the electronic pseudospin $\hat{\mathbf{S}}$ and the $^{53}$Cr nuclear spin $\hat{\mathbf{I}}_\text{Cr}=$3/2. The natural abundance of $^{53}$Cr is 9.501\%. The second term describes the $^{53}$Cr nuclear quadrupole interaction which is proportional to the electric-field gradient at the nuclear site. The nuclear quadruple interaction exists for any non-spherical nucleus with spin larger than 1/2. The third term describes the superhyperfine interaction of the pseudospin with the nuclear spins in the ligands, i.e., with $^{1}$H spins $I_\text{H}=$1/2 and with $^{13}$C spins $I_\text{C}=$1/2. The natural abundances of $^{13}$C and $^1$H are 1.07\% and 99.9885\%, respectively. The summation runs over all $^{13}$C and $^1$H atoms in the ligands. Note that the superhyperfine interaction represents the hyperfine coupling with nuclear spins of nuclei other than the one at which the electronic spin is localized. Therefore, in our case, the hyperfine coupling with carbon and hydrogen nuclear spins is referred to as superhyperfine interaction, while the hyperfine coupling with the Cr nuclear spin is simply denoted as a hyperfine interaction. In this work, we focus on quantification of the hyperfine and superhyperfine interactions and their effects on the $M_s$ sub-levels of the electronic ground state under different chemical environment by considering the {\bf M1}, {\bf M2}, and {\bf M3} molecules. Since we are interested in the hyperfine and superhyperfine interactions of the electronic ground spin-triplet state, we use the CASSCF(10,12) wavefunctions obtained from the state average over one root. 

\begin{table}[t!]
\centering
\caption{Isotropic Fermi contact $\mathbf{A}_\text{FC}^\text{Cr}$ and principal values of spin-dipole contributions $A_\text{SD,i}^\text{Cr}$ ($i=1,2,3$) of the $^{53}$Cr hyperfine matrix as well as the $^{53}$Cr nuclear quadrupole tensor $\mathbf{P}^\text{Cr}$ of the electronic ground state for all considered molecules. The coordinate system corresponds to the molecular magnetic anisotropy axes. Units of all values are MHz. For simplicity, we drop ''Cr'' in symbols below.}
\begin{tabular}{c|r|r|r}
\hline \hline
Properties      & \textbf{M1}  & \textbf{M2} & \textbf{M3}  \\
\hline
$A_\text{FC}$   &  94.9           &  95.8          &  95.4    \\
$A_\text{SD,1}$ & $-$0.27         &  0.4           &  0.3     \\
$A_\text{SD,2}$ &  0.14           & $-$0.4         &  $-$0.3  \\
$A_\text{SD,3}$ &  0.14           &  0.0           &  0.0     \\
$P_{xx}$        &  $-$1.5         &  0.8           &  $-$0.5  \\
$P_{xy}$        &  0.0            &  0.2           &  $-$0.1  \\
$P_{xz}$        &  0.0            &  0.7           &  $-$0.3  \\
$P_{yy}$        &  $-$1.5         &  0.4           &  $-$0.4  \\
$P_{yz}$        &  0.0            &  0.2           &  0.0     \\
$P_{zz}$        &  3.0            & $-$0.4         &  0.9     \\
\hline \hline
\end{tabular}
\label{tab:HF}
\end{table}

We discuss quantification of the $^{53}$Cr hyperfine interaction. Table~\ref{tab:HF} lists the calculated $^{53}$Cr hyperfine matrix $\mathbf{A}_\text{Cr}$ for all considered molecules. The $^{53}$Cr hyperfine coupling primarily originates from the Fermi contact term that ranges from 94.9 to 95.8 MHz for the three molecules. This term comes from significant spin density at the Cr nuclear site, which is caused by hybridization between unpaired Cr 4$s$ and 3$d$ orbitals. The spin-dipole contribution to the hyperfine coupling is less than 1\% of the Fermi contact term. The paramagnetic spin orbital (PSO) contribution is absent. We find that the signs of the hyperfine matrix elements are positive from a separate DFT calculation of the hyperfine matrix within the local density approximation\cite{Perdew1992} for the exchange-correlation functional without self-interaction correction using the {\tt FLOSIC} code\cite{FLOSIC_code,Pederson2014} (where the same basis sets as those of {\tt NRLMOL} are used). For all considered molecules, the elements of the $^{53}$Cr nuclear quadrupole tensor $\mathbf{P}_\text{Cr}$ turn out to be very small (0.1-1.5 MHz) and they reflect the molecular symmetries.

\begin{figure}[htb!]
\centering
\includegraphics[width=0.5\linewidth]{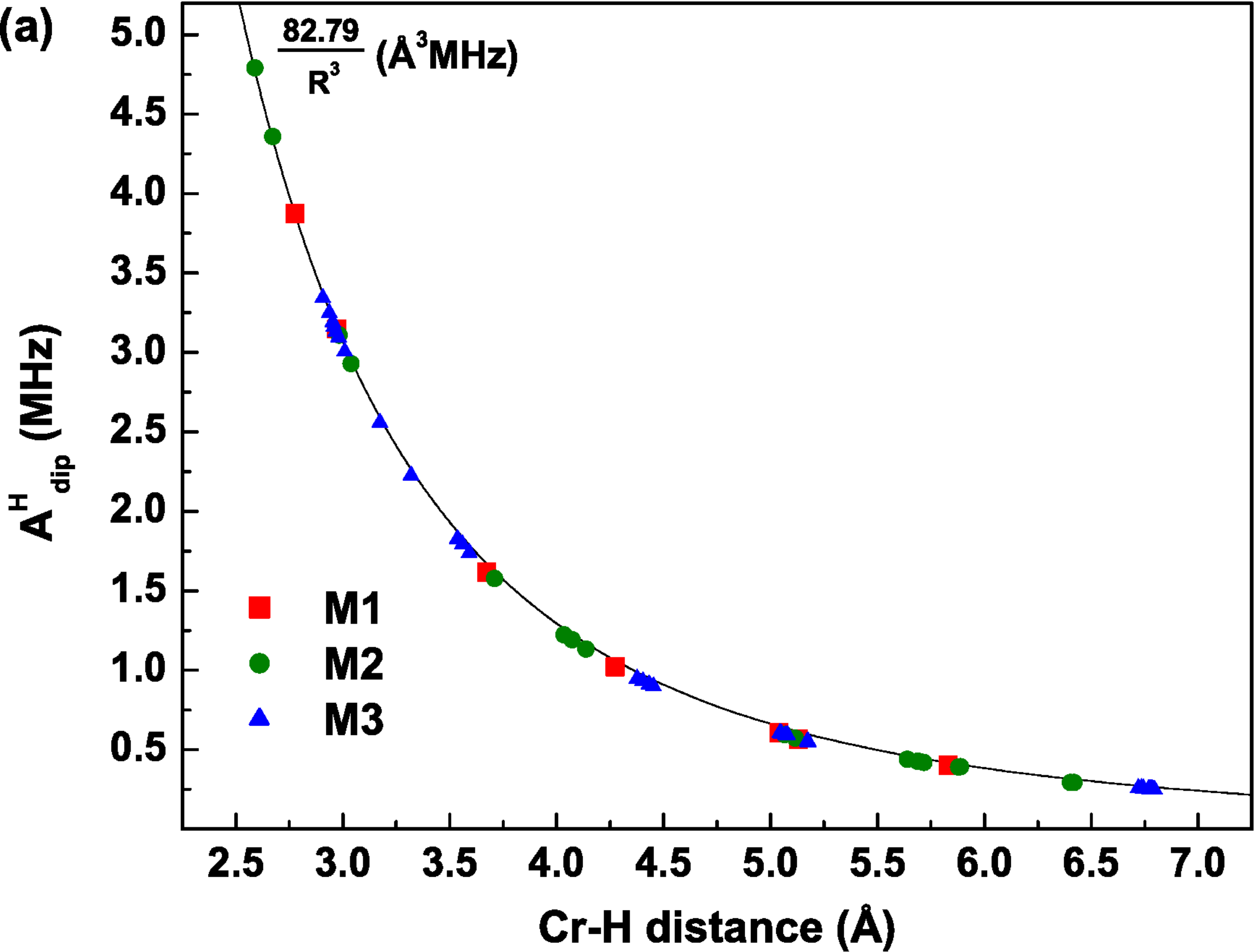}
\includegraphics[width=0.5\linewidth]{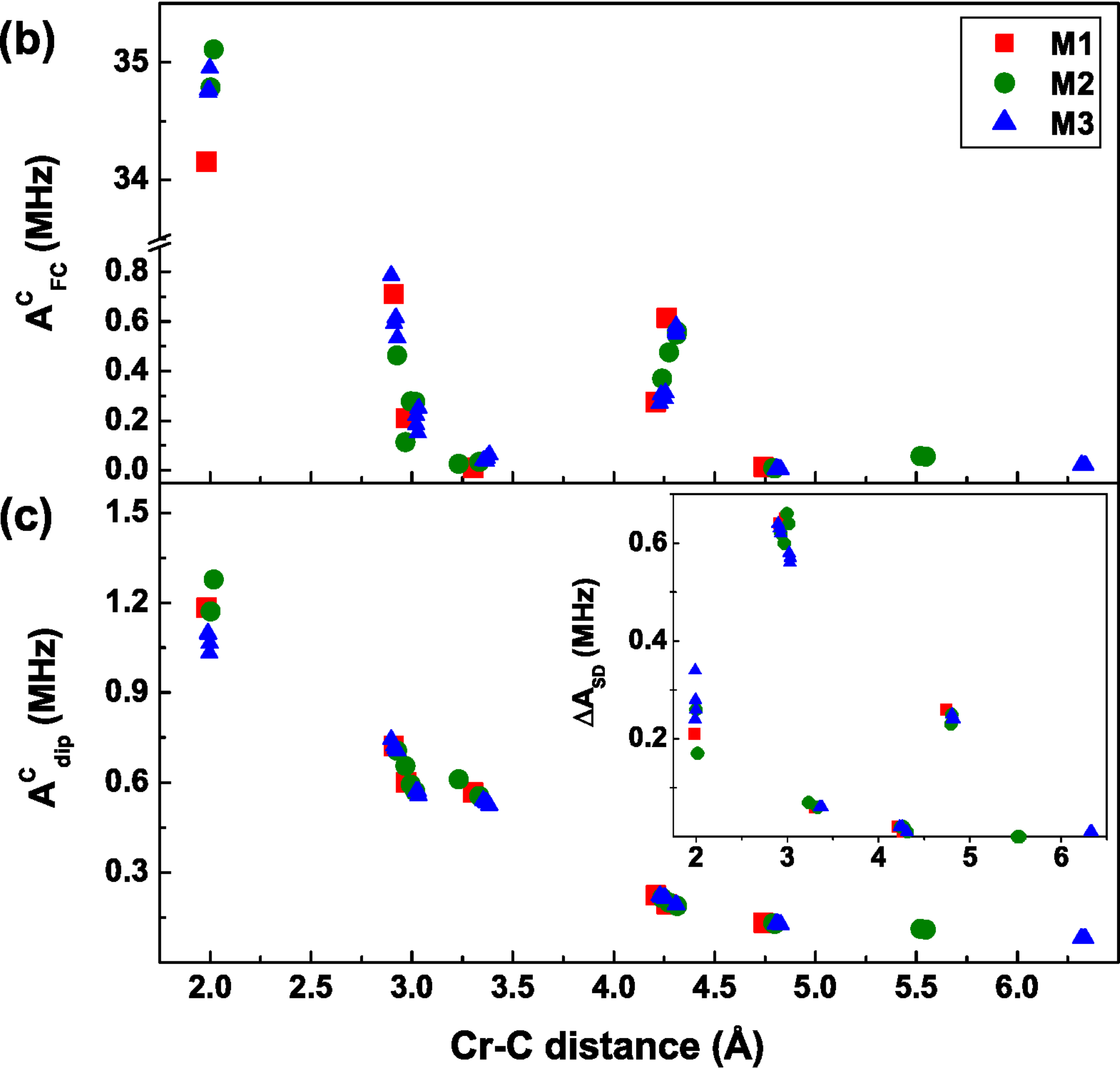}
\caption{(a) Spin-dipolar contributions to the superhyperfine matrix for the $^{1}$H nuclei as a function of the distance from the Cr ion for all considered molecules. (b) Fermi contact and (c) spin-dipolar contributions to the superhyperfine matrix for the $^{13}$C nuclei as a function of the distance from the Cr ion for all considered molecules. $A^{\text{H,C}}_\text{dip}=|A^{\text{H,C}}_\text{dip,1}-(A^{\text{H,C}}_\text{dip,2}+A^{\text{H,C}}_\text{dip,3})/2$|, where $A^{\text{H,C}}_\text{dip,i}$ ($i=$1,2,3) are three principal values of the spin-dipole contribution matrix to $\mathbf{A}_\text{H,C}$, and $A^{\text{H,C}}_\text{dip,1}$ is the largest-magnitude principal value. In the inset of (c), $\Delta A_\text{SD}=|A^{\text{C}}_\text{dip,2}-A^{\text{C}}_\text{dip,3}|$}.
\label{fig:SHF}
\end{figure}

The superhyperfine interactions of all the $^{1}$H nuclear spins mainly arise from the spin-dipole contribution which obeys the power law as a function of the nuclear distance $R$ from the Cr ion, as shown in Fig.~\ref{fig:SHF}(a). In this plot, $A^{\text{H}}_\text{dip}$ is defined to be $|A^{\text{H}}_\text{dip,1}-(A^{\text{H}}_\text{dip,2}+A^{\text{H}}_\text{dip,3})/2|$, where $A^{\text{H}}_\text{dip,i}$ ($i=$1,2,3) are three principal values of the spin-dipole contribution matrix to $\mathbf{A}_\text{H}$. Here $A^{\text{H}}_\text{dip,1}$ is set to be the largest-magnitude principal value. We observe that $A^{\text{H}}_\text{dip,2}$ is similar to $A^{\text{H}}_\text{dip,3}$. We find that the superhyperfine interactions of the $^{13}$C nuclear spins are more complex than the $^1$H isotope case, showing the following two features. Firstly, there is non-negligible spin density at the four nearest neighboring C sites. If these four C sites have nuclear spins, the dominant contributions come from the Fermi contact term of 34-35 MHz (Fig.~\ref{fig:SHF}(b)). Since the Fermi contact term decays exponentially with the distance from the nuclear site, for the $^{13}$C sites at intermediate distances from the Cr site, the superhyperfine interactions have contributions from both the Fermi contact and spin-dipole terms (Fig.~\ref{fig:SHF}(c)). Secondly, for the $^{13}$C sites far from the Cr site, although the spin-dipole contributions outweigh the Fermi contact terms, they do not follow the usual power law, $1/R^3$, and there are nonzero differences between $A^{\text{C}}_\text{dip,2}$ and $A^{\text{C}}_\text{dip,3}$.

\begin{figure}[htb!]
\centering
\includegraphics[width=1.0\linewidth]{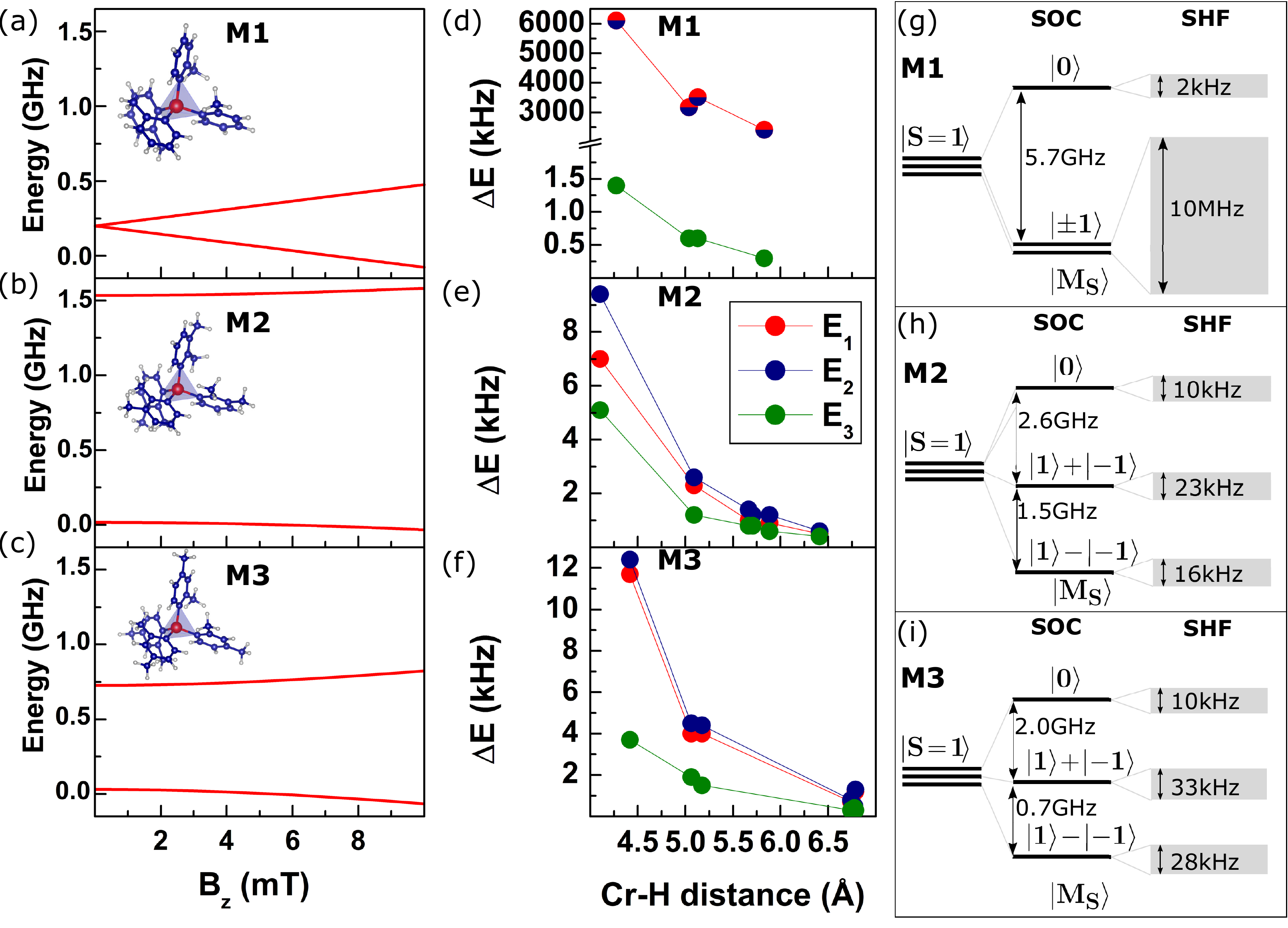}
\caption{(a)-(c) Magnetic field dependence of the $M_s=\pm1$ levels. (d)-(f) The broadening of the $M_s$ sub-levels of the electronic ground state due to superhyperfine interaction with a group of four $^{1}$H nuclear spins as a function of the average distance of the four H atoms to the Cr ion. Here, $E_3$ is a $M_s=0$ level while $E_1$ and $E_2$ denote, respectively, lower and higher $M_s=\pm1$ level. (g)-(i) Diagrams of the sub-level splitting and broadening of the electronic ground spin-triplet state due to the SOC and superhyperfine (SHF) coupling to the eight of $^{1}$H nuclear spins in the distance of $\sim$ 4-5~\AA~from the Cr ion. Grey bands represent the range of electronic-nuclear levels formed by the SHF interactions.}
\label{fig:SHyperH}
\end{figure}

The superhyperfine interactions can contribute to the decoherence of the electronic spin even at low temperatures.\cite{Takahashi2011,Lenz2017} The pairs of nuclear spins interacting with each other by magnetic dipolar forces may mutually switch their spin orientations at a negligibly low energy cost.\cite{Bloembergen1949} Such flip-flop processes induce  low-temperature fluctuations of the ligand nuclear spins, which is known as nuclear spin diffusion. Importantly, only the nuclear spins outside the so-called spin diffusion barrier\cite{Ramanathan2008} contribute to these fluctuations because the nuclear spins inside the barrier are strongly coupled to the electronic spin. The typical radius of the spin diffusion barrier in molecular systems is estimated to be 4-7 \AA~\cite{Zecevic1998,Graham2017}.~In particular, the (super)hyperfine couplings due to $^{53}$Cr and the four nearest neighbor $^{13}$C nuclear spins are, therefore, not expected to contribute to the decoherence. The nuclear spin diffusion leads to fluctuations of a local magnetic field seen by the Cr electron spin which, in turn, causes dynamic variations of the energy differences between $M_s$ sub-levels of the electronic ground state and influence the phase of the molecular spin state. Although rigorous studies of spin decoherence require simulations of spin dynamics,\cite{Seo2016,Lenz2017,Lunghi2019,JChen2020,Bayliss2022,Lunghi2022} the sensitivity to such decoherence effects can be, qualitatively, analyzed by inspecting a magnetic-field dependence of the two lowest $M_s$ sub-levels (Fig.~\ref{fig:SHyperH}(a)-(c)). These Zeeman diagrams are obtained using the effective spin Hamiltonian, Eq.~(\ref{eq:ZFS}), with an additional Zeeman term for the field along the $z$ axis. The two lowest zero-field split sub-levels of {\bf M2} and {\bf M3} are linear combinations of $M_s=+1$ and $M_s=-1$ sub-levels and so they are insensitive to an external magnetic field, to first order, which represents a clock transition.\cite{Shiddiq2016,Wolfowicz2013,Rubin2021,Bayliss2022,Collett2019} On the other hand, {\bf M1} has the $M_s=\pm1$ doublet and so there is a linear dependence on the magnetic field. Therefore, {\bf M2} and {\bf M3} are much more protected from the fluctuating ligand nuclear spins than {\bf M1}. Since the clock transition is induced by the ZFS $E$ parameter for integer spins, the insensitivity to an external or effective (fluctuating) magnetic field increases with increasing the $E$ value. Figure~\ref{fig:SHyperH} shows only the case of the magnetic field along the anisotropy ($z$) axis since the magnetic field dependence on the levels is much weaker for the field along the $x$ and $y$ axes, even in the absence of the clock transition.

The importance of the superhyperfine interactions can be also characterized by the broadening of the $M_s$ sub-levels of the electronic ground state due to coupling with the ligand nuclear spins. This broadening represents the splitting of each $M_s$ level into a group of close-lying electronic-nuclear levels as a result of the superhyperfine interaction. The width of such electronic-nuclear bands calculated using the Hamiltonian Eq.~(\ref{eq:ZFS}) and the superhyperfine term from Eq. (\ref{eq:SHF}), is plotted in Fig.~\ref{fig:SHyperH}(d)-(f) in the case of the superhyperfine interaction with groups of four $^{1}$H nuclear spins lying in the similar distances from the Cr ion as a function of Cr-H distance. Here, we only consider H atoms that are further than 4~\AA~from the Cr ion which roughly corresponds to a lower limit of the estimated spin diffusion barrier radius in magnetic molecules.\cite{Graham2017} In principle, the $^{13}$C nuclear spins outside the spin diffusion barrier could also contribute to decoherence. However, due to a very low natural abundance of this isotope, we expect that this contribution is much less important. Note that here we focus on an isolated molecule and intramolecular superhyperfine interactions. However, when the molecule is embedded in any type of environment, the coupling to nuclear spins outside the molecule can also play an important role. As expected, the broadening typically diminishes with increasing the Cr-H separation due to decreasing strength of the superhyperfine coupling. In order to obtain a rough estimate of the level broadening due to interactions with nuclear spins from outside the spin diffusion barrier, we consider superhyperfine interaction with eight of $^{1}$H nuclear spins in the distance of $\sim$ 4-5~\AA~from the Cr ion. The resulting level broadening is shown in Fig.~\ref{fig:SHyperH}(g)-(i). The crucial feature is that the broadening of the $M_s=\pm1$ is three orders of magnitude larger for the \textbf{M1} molecule ($\sim$ 10 MHz) than for the \textbf{M2} and \textbf{M3} molecules ($\sim$10 kHz). This is again a manifestation of the presence of the clock transition for the \textbf{M2} and \textbf{M3} molecules which strongly reduces the effect of superhyperfine interactions on the $M_s=\pm1$ levels. We can, therefore, speculate that the superhyperfine-induced decoherence may be much weaker for the \textbf{M2} and \textbf{M3} molecules than for the {\bf M1} molecule. However, in order to confirm this hypothesis, spin dynamics needs to be explicitly calculated.\cite{Seo2016,Lenz2017,JChen2020,Bayliss2022}

\begin{figure}[htb!]
\centering
\includegraphics[width=1.0\linewidth]{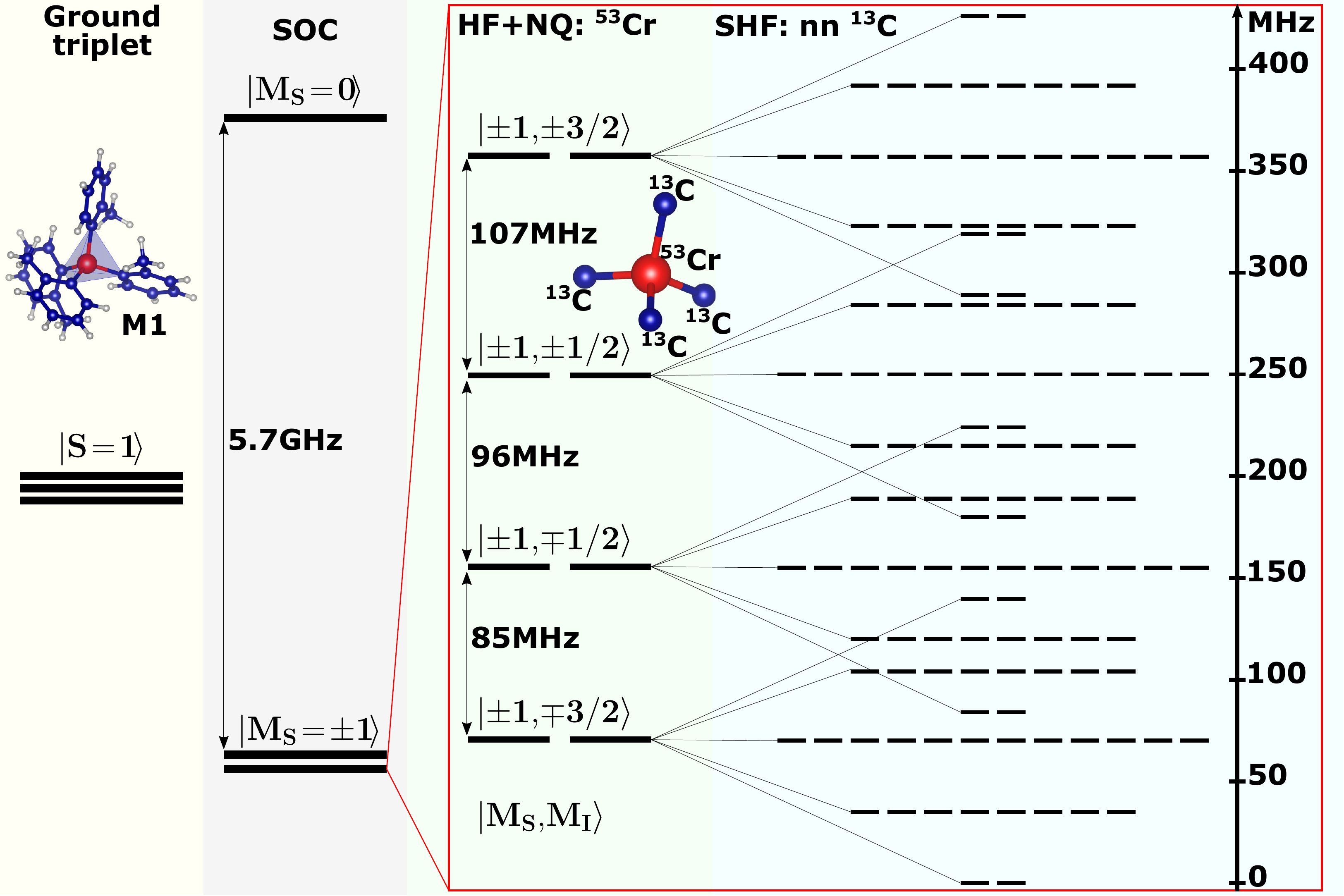}
\caption{Detailed level splitting of the spin triplet electronic ground state for the \textbf{M1} molecule which arises due to SOC, hyperfine and quadrupole couplings for the $^{53}$Cr nuclear spin, and the superhyperfine interaction with the four $^{13}$C nuclear spins closest to the Cr site.}
\label{fig:SHyperC}
\end{figure}

While the nuclear spin outside the spin diffusion barrier play an important role in decoherence, the nuclear spins inside the barrier that are strongly coupled to the electronic spin by hyperfine/superhyperfine interaction offer a promising platform for multiqubit registers.\cite{Hussain2018,Atzori2021,Ruskuc2022} Here, we consider the \textbf{M1} molecule and explore a potential quantum register formed by the electronic spin coupled to the $^{53}$Cr nuclear spin and $^{13}$C nuclear spins of four carbon atoms that are nearest neighbors to the Cr ion and strongly coupled to the electronic spin by the Fermi contact mechanism (see above). Note that while these nuclei (especially $^{13}$C) have a very low natural abundance, such nuclear configuration could be potentially realized by deliberate design with low-abundance isotopes. The electronic-nuclear spectrum corresponding to the spin triplet electronic ground state is calculated by diagonalizing the effective spin Hamiltonian that consists of Eqs.~(\ref{eq:ZFS}) and (\ref{eq:SHF}) terms and is shown in Fig.~\ref{fig:SHyperC}. Note that we neglected the nuclear spin-spin interactions as they are expected to be significantly smaller than the considered hyperfine and superhyperfine terms. Since the ZFS energy for {\bf M1}, 5.7~GHz, is much larger than the hyperfine/superhyperfine interaction energies ($\sim$100~MHz), the $M_s=0$ sub-level is well separated from the $M_s=\pm1$ sub-levels and we can focus on electronic-nuclear levels that originate from the $M_s=\pm1$ doublet. Let us first consider the coupling with the $^{53}$Cr nuclear spin. The positive signs of the hyperfine matrix elements suggest that the lowest energy levels are formed when the electronic and nuclear spin directions are antiparallel to each other such as $|M_s,M_I\rangle=|\pm1,\mp3/2\rangle$ levels, where $M_I$ is the projection of the $^{53}$Cr nuclear spin onto the magnetic easy axis. The small $^{53}$Cr nuclear quadrupole interaction is predominantly responsible for non-equidistant spacing of the electronic-nuclear levels $|M_s,M_I\rangle$. The consecutive electronic-nuclear level separation ranges from 85-107~MHz. With an addition of the superhyperfine coupling from the four nearest-neighboring $^{13}$C nuclear spins, the $|M_s,M_I\rangle$ levels are further split in a complex fashion (rightmost column in Fig.~\ref{fig:SHyperC}) and the overall broadening reaches about 425 MHz.

\section{Conclusions}

We investigate electronic spin-triplet and spin-singlet excitations and ZFS parameters of the ground spin-triplet state for a few Cr(IV)(aryl)$_4$ molecules with slightly different ligands and different molecular symmetries, using multireference {\it ab-initio} methods. Our calculated ZPL energies are in agreement with the experimental data. We show that high-energy spin-triplet and spin-singlet states play an important role in the ZFS parameters since SOC enters as a second-order effect and that SSC does not contribute to the ZFS parameters. The calculated ZFS parameters exhibit a significant dependence on ligands, which is consistent with the experimental data. We find that the uniaxial ZFS parameter of the ground state has a negative sign for all considered molecules. We suggest that the ISC mechanism can be used to achieve the optical spin initialization and that the efficiency of the spin initialization highly depends on ligand type. To acquire insight into electron spin decoherence and explore possible realizations of multiqubit registers, we calculate (super)hyperfine interactions of the $^{53}$Cr nuclear spin and $^1$H and $^{13}$C nuclear spins and study how these interactions induce the splitting of the electronic ground-state spin sub-levels. We show that the width of the sub-level splitting can be reduced by an order of magnitude when molecules have a significant ZFS $E$ parameter value, compared to molecules with $E=0$. We clarify that this effect is ascribed to a clock transition.

\begin{acknowledgement}
This work was funded by the Department of Energy (DOE) Basic Energy Sciences (BES) grant number DE-SC0018326. Computational support was provided by the Virginia Tech Advanced Research Center and the Extreme Science and Engineering Discovery Environment (XSEDE) under Project number DMR060009N which are supported by the National Science Foundation Grant number ACI-1548562. The authors are grateful to A. Karanovich for the hyperfine coupling calculation using the FLOSIC code.
\end{acknowledgement}

\begin{suppinfo}

The Supporting Information is available free of charge.
\\

DFT-relaxed atomic coordinates of the {\bf M1}, {\bf M2}, and {\bf M3} molecules; comparison of experimental and DFT-relaxed atomic coordinates; SA-CASSCF+CASPT2 spin-free energies for all considered molecules; DFT triplet-singlet excitations energies; analysis of the ZFS parameters as a function of number of spin-triplet and spin-singlet roots for {\bf M1}, {\bf M2}, and {\bf M3}; dominant configurations of the 13 spin-triplet and 15 spin-singlet roots for {\bf M1}; images of the active orbitals for the spin-triplet SA-CASSCF(10,12) calculations for ${\bf M1}$ with SA natural occupancies; ZFS parameters and f factors for {\bf M5}; SOC matrix elements; Racah parameters for {\bf M1} and {\bf M4}.

\end{suppinfo}

\bibliography{refs}

\end{document}